\documentstyle{article}

\font\elevenbf=cmbx10 scaled\magstep 1
\font\elevenrm=cmr10 scaled\magstep 1
\font\elevenit=cmti10 scaled\magstep 1

\font\ninerm=cmr9

\renewenvironment{thebibliography}[1]
 { \elevenrm
   \begin{list}{\arabic{enumi}.}
    {\usecounter{enumi} \setlength{\parsep}{0pt}
     \setlength{\itemsep}{3pt} \settowidth{\labelwidth}{#1.}
     \sloppy
    }}{\end{list}}

\newcommand{\dr}{\raise.3ex\hbox{$\stackrel{\leftarrow}{\partial
}$}{}}
\newcommand{\dl}{\raise.3ex\hbox{$\stackrel{\rightarrow}{\partial}$}{}
}
\newcommand{\eqn}[1]{Eq. (\ref{#1})}
\newcommand{\pl}{(\partial X^\mu )}
\newcommand{\plb}{(\partial X^\nu )}
\newcommand{\plg}{(\partial X^\rho )}

\newcommand{\bpl}{(\bar{\partial }X^\mu )}
\newcommand{\ds}{d_{\mu \nu \rho }}
\newcommand{\delb}{\bar \partial }
\newcommand{\na}{\nabla }
\newcommand{\ft}[2]{{\textstyle\frac{#1}{#2}}}

\newsavebox{\uuunit}
\sbox{\uuunit}
    {\setlength{\unitlength}{0.825em}
     \begin{picture}(0.6,0.7)
        \thinlines
        \put(0,0){\line(1,0){0.5}}
        \put(0.15,0){\line(0,1){0.7}}
        \put(0.35,0){\line(0,1){0.8}}
       \multiput(0.3,0.8)(-0.04,-0.02){12}{\rule{0.5pt}{0.5pt}}
     \end {picture}}

\renewcommand{\section}[1]{\addtocounter{section}{1}
\vglue 0.5cm
{\elevenbf\noindent \thesection . #1}
\vglue 0.4cm
\setcounter{subsection}{0}
}
\newcommand{\sectionsub}[1]{\addtocounter{section}{1}
\vglue 0.5cm
{\elevenbf\noindent \thesection . #1}
\vglue 0.2cm
  \setcounter{subsection}{0}}
\renewcommand{\subsection}[1]{\addtocounter{subsection}{1}
\vglue 0.2cm
{\elevenit \noindent \thesubsection . #1}
\vglue 0.1cm
 }
\begin{document}
\begin{titlepage}
\begin{flushright} KUL-TF-93/32\\ hepth@xxx/9307126  \\
                   July 1993\\
\end{flushright}

\vfill
\begin{center}
{\large\bf Regularization and the BV Formalism $^1$
}\\ \vskip 27.mm
{\bf  Walter Troost $^2$ and Antoine Van Proeyen $^{3}$
}\\ \vskip 1cm
Instituut voor Theoretische Fysica
        \\Katholieke Universiteit Leuven
        \\Celestijnenlaan 200D
        \\B--3001 Leuven, Belgium\\[0.3cm]
\end{center}
\vfill
\begin{center}
{\bf Abstract}
\end{center}
\begin{quote}
\small
We summarize the application of the field--antifield formalism to the
quantization of gauge field theories. After the gauge fixing, the main
issues are the regularization and anomalies. We illustrate this for
chiral $W_3$ gravity. We discuss also gauge theories with a non-local
action, induced by a matter system with an anomaly. As an illustration we
use induced WZW models, including a discussion of non--local
regularization, the importance of the measure and regularization of
multiple loops. In these theories antifields become propagating --- one
of many roles assigned to them in the BV formalism.

\vspace{2mm} \vfill \hrule width 3.cm
{\footnotesize
\noindent $^1$ presented at the conference "Strings '93", Berkeley, May
1993.  \\
\noindent $^2$ Bevoegdverklaard navorser, NFWO, Belgium;
E--mail : Walter\%tf\%fys@cc3.kuleuven.ac.be \\
\noindent $^3$ Onderzoeksleider, NFWO, Belgium;   E--mail :
FGBDA19@cc1.kuleuven.ac.be;
 }
\normalsize
\end{quote}
\end{titlepage}

\section{Introduction}
\baselineskip=14pt
\elevenrm

    During a dinner a few weeks ago, Mario Tonin reminded us that
Cardinal Bellarmino was right in the argument with Galileo. Bellarmino
agreed that indeed the picture of the planets going around the earth on
circles with epicycles could be simplified by putting the sun in the
center. But he insisted that there is not more {\elevenit truth} in the sun
being  the center of the universe rather than the earth.

    Ten minutes later, Pietro Fr\'e asked what he could expect from the
Batalin--Vilkovisky (BV) formalism, if he knows already the BRST
framework. We made the comparison with the above situation. One may learn
first the BRST formalism for usual Lie algebras. Then one may extend it
for the case of structure functions. Learning how to treat open algebras
is a further epicycle, and the next one may be the reducible algebras,
... . In this way, one does not need the BV formalism. But the latter
puts the sun in the center.

    We consider the BV formalism for Lagrangian gauge theories. Our main
subject will be, how to treat regularization in this framework, but we
will first recall the classical theory, in particular how to obtain, from
the classical action, a gauge--fixed action including ghosts. This text
is complementary to \cite{bvsb}. A longer text is in preparation
\cite{bvleuv}.

First we will indicate the main concepts: antifields, antibrackets, and
the extended classical action. This is suficient for the classical
Lagrangian theory. We will use chiral $W_3$ gravity \cite{HullW3chir} as
an example, of which a detailed treatment in the present context can be
found in \cite{anomw3}. Then we will show how the gauge fixing procedure
in this formalism just amounts to a canonical transformation
\cite{Warrenctr}. The fourth section will contain some generalities about
the quantum theory, and explain why we use mainly Pauli--Villars (PV)
regularization. Anomalies are the central issue of the fifth section.
There we will explain how a cohomological theorem leads to drastic
simplifications in the calculation of the general form (i.e. including
the antifield dependence) of anomalies. Also the cancellation of some
anomalies by background charges will be incorporated. In the final part,
we will see that there is another role for the antifields, when
considering anomaly--induced actions, as bona fide physical fields that
correspond to the additional degree of freedom (because the anomalies do
not cancel) of the quantum theory. We will illustrate this in
Wess--Zumino--Witten (WZW) models. We will treat multiple loops using
higher derivative regularization, and will also take into account the
issue of locality \cite{rawnlr}.

    Since our main example is $W_3$ gravity in the chiral gauge, let us
recapitulate the (well known) elements that we need (see
\cite{HullW3chir}). We will use only the realization in terms of free
bosonic fields. The classical fields are $\phi ^i=\{ X^\mu , h, B\}$,
and their action is
\begin{equation}
S^0=\int d^2x\left\{ -\ft{1}{2}\pl \bpl +\ft{1}{2}h\pl \pl + \ft{1}{3}B\ds
\pl \plb \plg\right\} \label{S0W3}
\end{equation}
with $h$ and $B$ Lagrange multipliers, imposing two constraints on the
classical level.
In order to close the algebra of these constraints, we have to impose
\begin{equation}
d_{\mu (\nu \rho }d_{\sigma )\tau  \mu }= \delta _{(\nu
\sigma }\delta _{\rho )\tau  }\ . \label{dsym}
\end{equation}
There are two gauge symmetries: the remnant of general coordinate and
dilatational symmetries, and the spin--3 symmetry. These imply that
the path integral
\begin{equation} Z(J_i)=\int {\cal D}\phi ^i\ \exp\left[ \frac{i}{\hbar
}S^0(\phi )+J_i\phi ^i\right]
\end{equation}
(where $\phi$ denotes $X,h$ and $B$) is not well defined,
some propagators are singular.
This problem is characteristic for gauge theories, and solved by gauge
fixing.

\section{Antifields, antibrackets, and the extended action}
The BV formalism for a Lagrangian gauge theory starts by introducing
a ghost field for every symmetry, and then an antifield for every field. This
leads to table~\ref{tbl:FieldsW3}.
\begin{table}[htf]\caption{Fields and antifields of chiral $W_3$ gravity}
\label{tbl:FieldsW3}
\begin{center}\begin{tabular}{||l|r|r||l|r|r||} \hline
   &stat&$gh$&      &stat&$gh$ \\ \hline
$X^\mu $&$+$ & 0  & $X^*_\mu $&$-$ &$-1$ \\
$B$&$+$ & 0  & $B^*$&$-$ &$-1$ \\
$h$&$+$ & 0  & $h^*$&$-$ &$-1$ \\
$c$&$-$ & 1  & $c^*$&$+$ &$-2$ \\
$u$&$-$ & 1  & $u^*$&$+$ &$-2$ \\ \hline
\end{tabular}\end{center}
\end{table}
Note that the ghosts are fermionic (statistics opposite to the symmetry),
and that the antifields have statistics opposite to that of the fields. The
classical fields have ghost number 0, and the ghosts have ghost number 1.
The ghost numbers of a field and its antifield always add up to $-1$.
In the space of functionals of fields and antifields, one introduces a
Poisson bracket--like structure (called antibrackets), where the antifields
take over the role of the canonical momenta.
Thus for example,
\begin{equation}
\left(X^\mu(x) ,X^*_\nu(y) \right)=\delta ^\mu _\nu\delta (x-y)
=-\left(X^*_\nu(x) ,X^\mu(y) \right)\ . \end{equation}
The new feature is that this antibracket is odd, i.e.
the Grassmann parity of $(A,B)$ is opposite to that of the product $AB$.
Equipped with this structure,
the BV formalism is analogous to the canonical formalism
of classical mechanics in many respects: brackets, canonical
transformations,\ldots . Note however that we will stay
within the Lagrangian framework, thus sticking to the covariant
formulation. For the Hamiltonian approach, we refer to the literature.

The extended classical action is a function of fields and antifields:
$S(\Phi ^A, \Phi ^*_A)$.
The part of the extended action which does not contain antifields is the
classical action:
\begin{equation}
S(\Phi^A ,\Phi_A ^*=0)= S^0(\phi ^i)\ .
\end{equation}
In our example it is  \eqn{S0W3}. This classical action
depends only on the classical fields $\phi ^i$, which in our example (with
$\mu $ running over $D$ values) means that it depends on $D+2$ fields. The
problem of singular propagators for the quantum theory can be restated as
\begin{equation}
rank\ \left( \partial _i\partial _j S^0\right) = D<D+2\ .
\end{equation}
To remedy this problem, we add a term
\begin{equation}
S^1= \phi ^*_i R^i{}_{a}c^a\ ,
\end{equation}
where $R^i{}_{a}$ is the variation of $\phi$ for the symmetry
transformation $a$. In $W_3$ this is \footnote{\ninerm\baselineskip=11pt
We omit $\int d^2x$, here and in the sequel, for actions and anomaly
expressions.}
\begin{eqnarray}
S^1&=& X _\mu ^*\left[\pl c+\ds \plb \plg u\right]\nonumber\\
& &+h^*\left[\na ^{(-1)}c+\pl \pl (u\partial B-B\partial u)
\right]+B^*\left[\na ^{(-2)}u-2B\partial c+c\partial B\right]
\end{eqnarray}
with $ \nabla  ^{(j)}=\bar{\partial }-h\partial -j(\partial h) $.
Note that for each symmetry this gives an extra contribution 2 to the
rank of $\partial _\alpha \partial _\beta S$. (To be more precise, we
look at this matrix at the surface where all fields of non--negative
ghost number vanish, and those of ghost number 0 satisfy the equations of
motion from $S^0$). In our example this contributes 4, and the total rank
is $D+4$, while introducing the 2 ghosts and all the antifields, raised
the dimension of this matrix to $2(D+4)$. The rank is thus half the
dimension, which turns out to be the maximum possible, due to the master
equation, \eqn{clmaster}. If the rank would still be lower, this would
signal that the transformations were dependent. This would be remedied by
introducing a term $c^*_a Z^a{}_{a_1}c^{a_1}$ where $Z$ stands for the
dependence relations between the symmetries and $c^{a_1}$ would be new
ghosts (so--called ghosts for ghosts, having ghost number~2).

The statement that $S^0$
is invariant under the transformations defined by $R^i{}_{a}$
is neatly expressed with the help of antibrackets: $(S^0,S^1)=0$.
This is a special consequence of imposing the more general relation
\begin{equation}
(S,S)=0  \label{clmaster}
\end{equation}
which contains, apart form the invariance of the action, also
the commutators, non--closure relations, (modified) Jacobi identities
and any other relevant data of the symmetry algebra. This relation is
called the (classical) master equation, and is a cornerstone of the
formalism. The extended action that satisfies this relation
will in general have, apart from $S^0$ and $S^1$, other pieces.
Also necessary are the following terms (written symbolically):
\begin{equation}
S^2=(-)^b\ft{1}{2}c^*_aT^a{}_{bc}c^cc^b
+(-)^{i+a}\ft{1}{4}\phi ^*_i\phi ^*_jE^{ji}{}_{ab}c^bc^a\ ,
\end{equation}
where $T$ are structure functions and $E$ appear in the non--closure
relations. For $W_3$ for example, the master equation is solved by
\begin{eqnarray}
S^2 &= &c^*\left[(\partial c)c+ (1-\alpha ) \pl \pl (\partial
u)u\right]+u^*\left[2(\partial c)u-c(\partial u)\right]\nonumber\\
& &-2\alpha  h^*(-2B\partial B^*-3B^*\partial B+\na^2h^*)(\partial
u)u - 2(\alpha+1)  X_\mu ^*h^*(\partial u)u\pl \ ,
\label{S2W3}\end{eqnarray}
where $\alpha $ is a free parameter. In general one can prove that
a solution (of ghost number 0) of the master equation always exists
(possibly containing terms with three or more antifields). This was proven
in \cite{BVJMP,henn} under some regularity conditions.
In \cite{bvleuv} it will be proven that there is always a {\elevenit local}
solution (local in space-time).
The essential ingredients and steps have been given already in
\cite{anomw3}. Also there, the regularity conditions for
this theorem have been weakened: previous versions were not applicable
to, e.g.,  $W_2$ (chiral gravity) and
chiral $W_3$. The theorem says moreover that the solution of the master
equation is unique up to a canonical transformation . In \cite{anomw3} it is
shown explicitly that indeed the parameter $\alpha$ in \eqn{S2W3} can be
removed by a canonical transformation.

For $W_3$ one finds that $S=S^0+S^1+S^2 $ is the full solution of the
master equation. It satisfies the three requirements: master equation,
properness (rank of the matrix of second derivatives is on--shell equal to
the number of fields) and classical limit (for vanishing antifields), which
were discussed also in section 3.1 of \cite{bvsb}.

In this section we used a basis where the fields have
$gh(\Phi )\geq 0$, while the antifields satisfy $gh(\Phi ^*)<0$. Such a
basis always exists, and will be called the `{\elevenit classical basis}'.
We define the antifield number ($afn$), as zero for the fields in this
basis, and equal to minus the ghost number for the antifields.
It is often convenient, as for example in the construction of the
extended action above, to work perturbatively in antifield number.

\section{Gauge fixing}
    The extended action has been constructed such that the rank of
$\partial _\alpha \partial _\beta S$ is half its dimension, i.e. equal to
the number of fields. Now imagine making a canonical transformation
between fields and antifields such that in the new basis (to be called
the {\elevenit gauge--fixed} basis) the submatrix $\partial _A\partial _B S$
(derivatives w.r.t. fields only) is invertible (on shell). The path
integral with the new fields as integration variables no longer suffers
from the gauge problem above: the propagators of the new `fields' have no
zero--modes. This canonical transformation is the gist of the gauge
fixing procedure in the BV framework.

In our example the canonical transformation needed is extremely simple:
we just have to interchange the names field and antifield  for $h$ and $B$
(like changing a momentum into a coordinate and vice versa in the
canonical formalism). Then $b\equiv h^*$ and $v\equiv B^*$
are fields and $b^*=-h$, $v^*=-B$ are antifields. The fields $\Phi ^A$ of
the gauge--fixed basis are then $\{X^\mu, v=B^*,b=h^*,c, u\}$.
For the  extended action this implies
\begin{equation}
S=-\ft{1}{2}\pl \bpl
+b\na ^{(-1)}c +v\na ^{(-2)}u
+\mbox{ antifield dependent terms}
\end{equation}
and in the antifield independent terms one recognizes the gauge fixed
action. This has no gauge symmetries left.

It may often be preferable (for example, to keep explicit Lorentz covariance)
to introduce extra fields to fix the gauge
(`non--minimal' solutions of
the master equation), and to perform the canonical transformation with the
help of a generating function called the gauge fermion $\Psi (\Phi )$:
\begin{equation}
\Phi '=\Phi \ ;\qquad  \Phi^*_A= \Phi '^*_A+\frac{\partial }{\partial \Phi
^A}\Psi \ .\end{equation}
For the examples we consider, this is a detour, which leads back to the
same gauge fixed action after eliminating auxiliary fields.

\sectionsub{Quantum theory}
\subsection{Measure and counterterms}
The quantum theory is not determined by the classical action alone.
First, there is the question what is the measure of the path integral, or
equivalently one needs a regularization. Secondly, what is the quantum
action? The (extended) action $S$ may have terms of order $\hbar $.
For now, we will use the following expansion (later we will see that
half-integer powers of $\hbar$ may also be needed):
\[ W=S+\hbar M_1 +\hbar ^2M_2 +...\,\,,
\]
where $M_1$, ... are integrals of local functionals.
The extra terms may have infinite coefficients  related to the
regularization, or may be
finite. The `counterterms' $M_i$ are restricted by extra
requirements, e.g. rigid symmetries,
renormalization conditions, or experimental numbers.
Symmetry usually fixes the structure of these terms, leaving only
some undetermined constants.

The two questions above are connected. A change in the regularization
can be compensated by a change of the $M_i$. E.g. the regularization
scheme which will be introduced below depends on a mass matrix, and there
is a formula \cite{anombv} giving the change of $M_i$ which is connected
to a change of the mass matrix.

\subsection{Regularization procedure}
The quantum theory starts from the path integral
\[ Z(J,\Phi ^*)= \int {\cal D}\Phi \ \exp\left[ \frac{i}{\hbar }S(\Phi,\Phi
^* )+J\Phi \right]  \ .\]
For a variety of reasons it is advisable not to put the antifields equal
to zero (or, as is usually done, to put
$\Phi^*=\partial\Psi/\partial\Phi$, which is the same as putting them to
zero after a gauge--fixing canonical transformation). We recall that this
antifield dependence was already used, in the form of dependence on
sources for the BRST transformations, in \cite{ZinnJustin}, to study the
renormalization properties of ordinary non--abelian gauge theories (where
$S$ is linear in $\Phi^*$). For anomalous theories this dependence will
describe additional degrees of freedom.

    For the regularized theory, the path integral is considered to be the
limit when some cutoff $M$ is removed from  a well--defined path integral,
$Z_R = \lim_{M\rightarrow \infty }Z_M\ $.
One might read this $M$ as $1/(n-4)$ in dimensional
regularisation, but it is hard to see how to make sense of  $Z_M$ as a
path integral based on an action
for $n$ not an integer. Another proposal would be a
point--splitting or lattice regularization, but as we
try to work in the context of local field theories, we prefer not to use
these. We will use a two-step procedure, first modifying propagators
with higher derivative terms, and then applying
the method of Pauli and Villars (PV).
For the present formalism, both steps have the advantage that they can be
incorporated in a Lagrangian framework. The higher derivatives serve to
regularize all the diagrams apart from the one--loop contributions, while
PV should regularize the remaining one--loop divergences.
When we only consider the quantum corrections of order $\hbar $ (one--loop
diagrams), then PV  alone suffices.

\subsection{PV action}
To determine the PV action, we introduce for every field $\Phi ^A$, another
field $\chi ^A$ as PV--partner, with statistics such that a minus
sign is introduced in each  loop. For a more detailed description, see
section 5.1 of \cite{bvsb}. Each PV-field comes with its own antifield.
We will collectively denote the original fields and antifields as
$z^\alpha$, the PV partners as $w^\alpha$:
$ z^\alpha = \{ \Phi ^A, \Phi ^*_A\}\ ; \  w^\alpha =\{ \chi
^A,\chi ^*_A\} $.
Then suppose that we have found an action with higher derivatives which
regularizes the higher loops. If $S_\Lambda $ is this extended action,
(when considering only one loop, one may replace
$S_\Lambda $ with $S$), then we propose for the full action
\begin{equation}
 S_M= S_\Lambda (z) + w^\alpha \left(
\frac{\dl}{\partial z^{\phantom{\beta }\!\!\!\alpha}
}S_\Lambda \frac{\dr}{\partial z^\beta } \right) w^\beta -\ft12 M^2 \chi ^A
T_{AB} \chi ^B\ . \label{PVhdact}
 \end{equation}
Here, $T_{AB}$ is an matrix that is arbitrary but invertible for the
propagating fields. Note that in the mass term only the PV fields, and not
the PV antifields appear. On the other hand, $T$ may depend on ordinary
fields and antifields. The general prescription given here is certainly
not the most general one, but it satisfies some nice properties: to
one--loop order it commutes with canonical transformations, satisfies the
master equation automatically, and corresponds to the addition
(classically) of a cohomologically trivial system when
$M\rightarrow\infty$.

\sectionsub{Anomalies}
\subsection{General theory}
Since we keep the antifield dependence, all the usual generating
functionals will generically depend on $\Phi^*$ also. The one--particle
irreducible functional, defined as usual, is then a functional both of
some classical fields $\Phi_{cl}$ and on $\Phi^*$. The Ward identities,
as derived by Zinn--Justin \cite{ZinnJustin} for the usual non--abelian
gauge theories, find a very natural expression in this framework. The
general formal derivation is simply a matter of filling in the
definitions and doing a partial integration:
\begin{equation}
 (\Gamma ,\Gamma ) =\Delta \cdot \Gamma\ , \label{ZJME}
\end{equation}
where $\Gamma$ is the effective action, and the antibracket is now with
respect to $\{\Phi_{cl},\Phi^*\}$.
The right hand side symbolizes an operator insertion which computes the
eventual anomalies, its path integral expression is
\[ (\Delta \cdot \Gamma) =\frac{-2i\hbar}{Z}\int{\cal D}\Phi\ {\cal
A}(\Phi ,\Phi ^*) \ \exp \frac{i}{\hbar }(W+J\Phi )\ .\]
Using formal manipulations on the path integral, the anomaly
${\cal A}(\Phi ,\Phi ^*)$ takes the form
\begin{equation}
 {\cal A}=\Delta W +\frac{i}{2\hbar }(W,W) \ ,\label{cAform}
\end{equation}
where
\begin{equation}
\Delta =\frac{\partial }{\partial \Phi ^A}
\frac{\partial }{\partial \Phi _A^{\phantom{A}\!\!\!*}}\ .\label{defDel}
\end{equation}
However, for a local action $\Delta S$ gives an expression proportional
to $\delta (0)$, and is thus ill--defined, indicating where regularization
necessarily enters. Continuing anyway formally for a moment, we
expand ${\cal A}=0$ in $\hbar $, obtaining the equations
\[ (S,S) =0 \ ;\qquad \Delta S +i (S,M_1)=0\ ;\ \ldots \ .\]

Now we consider this in a regularized version, using the PV framework
and including only single  loops.
For the regularized path integral we take
\begin{equation}
 Z_R(J,\Phi ^*)
= \exp -\frac{i}{\hbar }W_c(J,\Phi ^*)=
 \lim_{M\rightarrow \infty }\int {\cal D}\Phi {\cal
D}\chi \ \left.\exp\left[ \frac{i}{\hbar }W_M(z,w)+J\Phi
\right]\right|_{\chi ^*=0} \ ,
\end{equation}
and $\Gamma$ introduced above is of course the Legendre transform of $W_c$.
The measure we define so that
$ \Delta W_M $ vanishes (up to terms quadratic in PV (anti)fields).
 Indeed, the contributions of the ordinary fields and the PV fields in
\eqn{defDel} cancel (again see section 5.1 of \cite{bvsb} for more details).
The anomaly arises from the fact that $(S_M,S_M)\neq 0$ generically,
due to the non--invariance of the PV mass term. The integral over the
PV fields gives for the
terms quadratic in PV fields a propagator, which is of order $\hbar$, and
we obtain
\begin{equation}
{\cal A}= Tr \left[J\, \exp ({\cal R}/M^2)\right] +i(S,M_1)+{\cal
O}(\hbar )\ .\label{cAreg}
\end{equation}
where $J$ is like the jacobian matrix of the BRST transformations and
${\cal R}$ is a regulator constructed from the action. The full expressions
are \cite{measure}
\[ J^A{}_{B}= \frac{\dl }{\partial \Phi _A^{\phantom{B}\!\!\!*}}S
\frac{\dr}{\partial \Phi ^B}+\frac{1}{2}(T^{-1})^{AC}\left(
T_{CB},S\right) (-)^B\
;\qquad {\cal R}^A{}_{B}=(T^{-1})^{AC}\left( \frac{\dl}{\partial \Phi ^C}S
\frac{\dr}{\partial \Phi ^B}\right) \ .
\]
Comparing \eqn{cAreg} with \eqn{cAform}, we have effectively obtained a
one--loop regularized
definition of $\Delta S$. Its value depends on the matrix $T$.
Different choices of $T$ give values of $\Delta S$ which differ
by $(S,G)$ for some local function $G$.
In practice, Gilkey's
formulas for the heat kernel \cite{heatk,bvleuv} are very useful.

Both for the formal expression and for the regularized one, one can
prove that
\begin{equation}
(S,{\cal A})=0\ .\label{WZcc}
\end{equation}
This is the expression in BV language of the Wess--Zumino consistency
conditions. Thus we automatically obtain  the `consistent' anomalies.
This was to be expected, since these conditions are satisfied for finite
values of $M$ also.

\subsection{Simplifications using a theorem on antibracket cohomology}
The calculations of the one--loop anomalies for chiral $W_3$
has been greatly simplified \cite{anomw3} by using  a theorem on the
cohomology of the nilpotent operator ${\cal S}F\equiv (F,S)$.
Whereas in \cite{henn} this was discussed in general terms, in
\cite{anomw3} it has been shown to be applicable also within the
set of local functions.

The  theorem (see \cite{anomw3}, theorem 3.1) states that the cohomology
of ${\cal S}$ in the set of
local functions of fields and antifields is equivalent to the
`weak' cohomology of an operator $D^0$ which acts in the set of local
functions of fields (or antifields) with non--negative ghost number.
Thus it is  most useful in
the classical basis, where $D^0$ acts on fields only:
$D^0 F^0 = \left.( F^0,S)\right|_{\Phi ^*=0}$.  `Weak'
cohomology means that field equations of $S^0$ (i.e. the part of the action
which depends on fields of ghost number 0) may be used freely.

The theorem implies that it is sufficient to calculate the part of the
possible anomaly which depends only on the fields of
the classical basis, call it ${\cal  A}^0$. Given ${\cal  A}^0$, the WZ
consistency condition \eqn{WZcc} implies that ${\cal A}$ is then fixed up
to terms $(S,M)$, with $M$ the integral of a local function. Such terms
are removable anyway by including $M$ as a local counterterm.

This way, the same results are obtained as in \cite{W3cal}. Removing all
dependence on the antifields of the classical basis (such as the antighosts
$b=h^*$ and $v=B^*$) one finds
\begin{eqnarray*} {\cal A}^0&\approx& -\frac{1}{24\pi } \left( c\delta
^\mu _\nu +2u\ds \plg\right) \partial ^3 \left( b^*\delta _{\mu \nu
}+2d_{\mu \nu \rho }v^*\partial X^\rho\right) \\ &&+\frac{100}{24\pi }\,
c\, \partial ^3 v^* +\frac{\kappa }{2\pi } (u\partial v^*-v^*\partial
u)(\partial ^3X^\mu )\pl\ , \end{eqnarray*}
where $\approx$ means `up to field equations'(the complete result is in
\cite{anomw3}).
We call attention to the fact that the anomaly depends on $b^*$ and
$v^*$, which are antifields in the gauge--fixed basis. In other
treatments one has to include background fields in the gauge fixing to
see the anomalies (for example, in the gauge $h=0$ the anomaly would not
show up). In the BV formalism, this task is taken over by the antifields,
without such extra ingredients.

\subsection{Background charges}
It is well known that the anomalies of the chiral $W_3$ model of \eqn{S0W3}
can be cancelled
by introducing background charges.  In our framework these correspond to
terms with
$\sqrt{\hbar }$ in the quantum action, which takes the form
\begin{equation}
W=S+\sqrt{\hbar}\,M_{1/2}+\hbar\, M_1+\ldots \ .
\label{Wh12} \end{equation}
Then the master equation ${\cal A}=0$ expands as
\begin{equation} (S, S)=(S, M_{1/2})=0\ ;\qquad
 (S, M_1)=i\Delta S-\frac{1}{2}(M_{1/2}, M_{1/2})\ .
\label{mem3} \end{equation}
In view of the cohomology theorem, $M_{1/2}$ is fixed when the terms
independent of the antifields (of the classical basis) are given.
These are
\[ M_{1/2}{}^0=a_\mu h(\partial ^2X^\mu )+e_{\mu \nu }B\pl \plb\ , \]
where
$a_\mu $ and $e_{\mu \nu }$ are constants (background charges). The
theorem says that $(S, M_{1/2})=0$ has a solution
for $M_{1/2}$ if and only if $D^0M_{1/2}\approx 0$. This gives some
conditions on the background charges \cite{Romans}. Taking into account
\eqn{dsym}, there is a unique solution for $e_{\mu\nu}$ for each value
of $D$ (the range $\mu$), and $a_\mu$ is still arbitrary. The latter
is also fixed by requiring that \eqn{mem3} has a solution for $M_1$, so
that we have an anomaly--free action of the form \eqn{Wh12}.
The simplification brought about by  the theorem is that the analysis
could be performed in the restricted space of fields of non--negative
ghost number.

We have used PV regularization for one loop, to be combined
with higher derivatives for more loops as in the next section.
The reason is that these regularizations are possible at the
level of the Lagrangian, thus leaving the BV--setup intact. If the
regularization does not break symmetries, then there are no anomalies. In
the usual cases the higher derivatives can be taken to be covariant, so
that anomalies only come from the PV mass term, and are genuine one--loop
anomalies. For $W_3$ it is not known how to define fully covariant
derivatives (with a finite number of terms) and this clarifies the
existence of anomalies beyond 1 loop in these theories.

\section{Induced theories}
Now we will consider theories where the anomalies are not cancelled, i.e.
$ (\Gamma , \Gamma )\neq 0\ , $ implying that
$\Gamma (\Phi _{cl}, \Phi ^*)$ describes new quantum degrees of freedom
in addition to the classical ones. They manifest themselves as
propagating antifields. For conformal matter coupled to 2--d gravity,
there is generically an anomaly
proportional to ${\cal A}=c\, \partial ^3 h$ .
The $h$--dependence of effective action obeying \eqn{ZJME} is proportional
to the non--local expression:
\[ \Gamma \sim \partial ^2 h\frac{1}{\partial \bar
\partial -\partial h\partial }\partial ^2h\ .\]
It was noticed by Polyakov that this is local in $f$, where
$ h= \frac{\bar \partial f}{\partial f}\ $.
This fact can be understood \cite{W2WZ} by first
 expressing  the original action in $f$ and performing
the regularization using as a mass term
$ M^2 X^\mu X^\mu (\partial f)$.
This mass term is invariant under
$\delta X=\epsilon \partial X,\delta f=\epsilon \partial f$, there is no
anomaly and zero induced action for $h$. However, we really wanted a
regularisation that is local {\elevenit in} $h$, not just in $f$, which
can be done using the mass term $M^2 X^\mu X^\mu$. This is just a change
of regularization, and, as mentioned before, this amounts to changing the
theory with a local counterterm, local in $f$ that is. Actually, there is
another integral left, over a parameter interpolating between the
regularizations, but this is trivial here. Therefore, the final induced
action is local in $f$.

A similar reasoning can be given for the WZW model, to which we now turn.
This will again illustrate some  aspects of the quantization procedure,
like the importance of the definition of the measure, non--local
regularization including also higher loops\cite{rawnlr}.
This model was also discussed, from a different point of view,
at this conference by P. van Nieuwenhuizen \cite{PvNWZW}.
The WZW action  arises as an induced action from
\begin{equation}
S= \psi^t (\bar \partial
-A)\psi -\psi ^* c\psi + A^*\bar D(A)c -c^* c c
\ ,\label{psitoWZW}
\end{equation}
with $\bar D(A)c\equiv \bar  \partial c - [A ,c]$,
where the fields $A$ and $c$ are in the adjoint representation of a
Lie--algebra, and $\psi$ is in an arbitrary representation.
The gauge fixing
of this model can be done by considering $A^*$ as a field (antighost),
and $A$ is then an antifield. Integrating out the fermions, the anomaly
implies that the resulting
action is a non--local action for the `antifield' $A$:
\begin{equation}
\Gamma[A] = \frac{k}{2\pi x}   \left\{ \,\,
\frac{1}{2} \, A \,  \frac{\partial}{\bar\partial}A \, - \frac{1}{3} \,
\frac{1}{\bar\partial} A \, \left[ A, \frac{\partial}{\bar\partial} A
\right]+\ldots \right\}  \label{GamA}
\end{equation}
Again it is possible to compare different regularizations \cite{W2WZ}
to arrive at the
conclusion that this action is (almost) local  when written in terms of a
group element $g$. Almost, since this time the integral over the
interpolating parameter remains (as the third space coordinate below):
\begin{equation}
 \Gamma [A=\delb g g^{-1}]=k\Gamma ^0[A]=-k\ S^+(g)\ ,
\end{equation}
with
\begin{equation}
S^+ (g)=  \frac{1}{4\pi  x}\left(  \int d^2 x \;
 \left\{ \partial  g^{-1} \delb g \right\}
+ \frac{1}{3} \int d^3 x\; \epsilon ^{\alpha \beta \gamma } \,
 \left\{ g_{,\alpha } g^{-1} g_{,\beta }
g^{-1} g_{,\gamma } g^{-1} \right\}\right) \ .
\end{equation}
To investigate the antifield degree of freedom, one takes the action
\eqn{GamA} as a starting point and quantizes. The generating functional
of interest is then
\begin{equation}
Z[u]=e^{\displaystyle -W[u]}=\int {\cal D} A \, e^{\displaystyle
-\Gamma[A]+ \frac{1}{2\pi x}
    \left\{ u\, A \right\}  } \ . \label{defw}
\end{equation}
It can be argued in different ways (\cite{frWZW,kpz},
see \cite{rawnlr} for a review) that
it enjoys the remarkable renormalization property
\begin{equation}
W[u]=Z_W W^0[Z_u u]\ ,
\end{equation}
where the Legendre transform of $\Gamma^0[A]$ defines $W^0[u_0]$:
\begin{equation}
W^0[u]=\min_{\{ A\} }\left( \Gamma^0[A]-\frac{1}{2\pi x}
 \{ u\,A\} \right)\ .
\end{equation}
Different formal arguments lead to $Z_W=k+2 \tilde h$, but
the values obtained for $Z_u$ vary. We will discuss now a regularized
argument, which makes use of the antifields in the regularization.
To start \cite{kpz}, consider the non-chiral induced action depending on
the gauge fields $A$ and $\bar A$:
\begin{equation}
\Gamma ^v[A,\bar A]= \Gamma  [A]+\overline{\Gamma }[\bar A]-\frac{k}{2\pi x}
\left\{A\,\bar A\right\}\ . \label{nceffact}
\end{equation}
It can be viewed as arising from two chirally conjugated matter systems
like \eqn{psitoWZW}.
If one separately regulates both chiral parts (a separate mass term for
each chiral fermion) then one obtains of course
just the first two terms in \eqn{nceffact}. Regulating in a vector
invariant way (a vector
invariant mass term for the fermions), the result differs from the
previous one by a finite local counterterm, which is the third term in
\eqn{nceffact}.
This term can be computed explicitly, example 4.1 in \cite{anombv}
does it in 4 dimensions.
To quantize this action, one has to take the
vector invariance into account, with
corresponding ghost $c$, and furthermore the antifields $A^*$, $\bar
A^*$ and $c^*$. The gauge fixing is done by considering $\bar A$ as
an antifield, and $\bar A^*$ as a field. We thus consider the path
integral
\begin{equation}
Z_{nc}(\bar A)\equiv\int {\cal D}A\,{\cal D}\bar A^*\, {\cal D}c\
e^{\displaystyle \Gamma^v(A,\bar A)+\bar A^* D(\bar A) c}\ .
\label{ncpathint} \end{equation}
For our present purpose it is sufficient to take  this path integral at
$A^*=c^*=0$, and not to introduce
sources for the fields $A$, $\bar A^*$ and $c$.
The vector gauge invariance allows us to fix $\bar A$
 to an arbitrary value, and if there is no anomaly then $Z_{nc}$ will not
 depend on it. Thus by investigating the vector anomaly we obtain the
dependence on $\bar A$ (an antifield from the present point of view).
For a one--loop calculation, we introduce PV fields with the
mass terms \cite{rawnlr}
\[ M^2 \underline{A} \frac{1}{\delb^2} \underline{A} + M
\underline{\bar A^*} \underline{c} \ ,
\]
where the underscore denotes PV--partners.
These mass terms are not invariant, but the anomalies cancel.
In \cite{rawnlr}
also more loops were considered. In the spirit of the method of
higher derivative regularization, the added term was taken to be the
vector--covariant
\begin{equation}
\frac{1}{\Lambda^2}\left(\partial A-\bar\partial\bar A+[A,\bar
A]\right)^2\ .
\end{equation}
Note the $\bar A$--dependence, which from the present point of view is
really an antifield--dependence.
This term regularizes all but the one--loop diagrams. Then one
could add the PV sector as in \eqn{PVhdact}, but it is shown in
\cite{rawnlr}
that a simpler action regularizes all the diagrams.
It turns out that also for multiple loops there is no
anomaly, and thus \eqn{ncpathint} is independent of $\bar A$.
One can therefore normalize $Z_{nc}(\bar A)=1$. We can now compare
\eqn{ncpathint} with \eqn{defw}. Identifying
$\bar A$ with the source $u$,
\begin{equation}
W[u]=(k+2\tilde h)W^0\left[\frac{1}{k}u\right]\ .\label{Wvecinv}
\end{equation}
The above amounts to a specific method to fix what we mean by the
functional integral measure ${\cal D}A$.
The result for the field renormalization factor is valid for this specific
regularization, but not necessarily for others.
To develop some understanding for other approaches, one may relate this
measure to an integration measure over group variables by the usual
formula:
\begin{equation}
{\cal D} A = {\cal D} g \; e^{\displaystyle  2\tilde{h}S^+[g]} \ .
\end{equation}
Checking the consequences for this ${\cal D}g$, one finds that it
is not Haar invariant:
\begin{equation}
\left[{\cal D}(hg) (hg)^{-1}\right]\neq\left[{\cal D}g\ g^{-1}\right]\ .
\end{equation}
It differs from the Haar invariant measure only by a counterterm that is
local in $g$ however. If one uses that Haar invariant measure, then for
the new path integrals one finds
\[ W_{\rm Haar}(u)=(k+2\tilde h)W^0\left(\frac{1}{k+2\tilde h }u\right)\ ,\]
differing from \eqn{Wvecinv}  in the field renormalization factor.
This shows once more the importance of the definition of the measure.

\section{Conclusions}
    We have seen that the BV formalism unifies many aspects of the
quantization in one formalism. It is not just that it offers the
possibility to treat all sorts of gauge theories, but also it phrases
Ward identities, WZ consistency conditions, induced actions, ... in a
very natural language. Another indication that the sun is in the center
is the multiple role played by the antifields, viz. first to generate the
(Faddeev--Popov) antighosts, then also as sources for BRST
transformations, as background fields in general gauges, and finally to
describe propagating quantum degrees of freedom appearing in induced
actions arising from antifield--dependent effective actions.

\section{Acknowledgments}
We thank the organizers of the conference for providing a stimulating
atmosphere. Also discussions with Joaquim Gom\`{\i}s, Alexander Sevrin,
Ruud Siebelink and Stefan Vandoren are gratefully acknowledged.

\end{document}